\documentclass[a4paper]{article}

\usepackage{graphicx}
\usepackage{color}
\usepackage{graphicx}
\usepackage{txfonts}

\newcommand{\Mpc}{$h^{-1}$\thinspace Mpc}

\newcommand{\vmh}{h^{-1}\mathrm{Mpc} }

\newcommand{\angstrom}{\mbox{\normalfont\AA}}

\DeclareMathAlphabet{\pazocal}{OMS}{zplm}{m}{n}

\begin{document}

{\begin{center} COLLAPSE, CONNECTIVITY, AND GALAXY POPULATIONS\\
    IN SUPERCLUSTER COCOONS: THE CASE OF A2142\footnote{Paper presented at the Fourth Zeldovich 
meeting, an international conference in honor of Ya. B. Zeldovich held in Minsk, Belarus on September 7--11, 2020. Published by the 
recommendation of the special editors: S. Ya. Kilin, R. Ruffini and G. V. Vereshchagin.}
\end{center}}

\begin{center} {\bf Maret Einasto} \end{center} 

{\em Tartu Observatory, Tartu University, Observatooriumi 1, 61602 T\~oravere, Estonia}\\

\abstract {
The largest galaxy systems in the cosmic web are superclusters, 
overdensity regions of galaxies, groups, clusters, and filaments. 
Low-density regions around superclusters are
called basins of attraction or cocoons. 
In my talk I discuss the properties 
of galaxies, groups, and filaments in the A2142 supercluster and its cocoon
at redshift $z \approx 0.09$.
Cocoon boundaries are determined by the lowest density
regions around the supercluster. We analyse the 
structure, dynamical state, connectivity, and galaxy content of the 
supercluster, and its high density core with the cluster A2142. 
We show that the main body of the supercluster is collapsing, and long filaments
which surround the supercluster are  detached from it.
Galaxies with very old stellar populations lie not only in the central 
parts of clusters and groups in the supercluster, but also in the poorest
groups in the cocoon. 
}


\onecolumn
\nopagebreak[4]

\section{Introduction}

The large-scale distribution of galaxies can be described as the  
supercluster-void network or the cosmic web. The cosmic web consists  
of galaxies, galaxy groups, and clusters connected by filaments
and separated by voids \cite{1}.
The largest systems in the cosmic web 
are galaxy superclusters, defined as the high-density regions in
the cosmic web. Superclusters act as great attractors, growing through the inflow of matter 
from surrounding low-density regions. Therefore superclusters
have also been defined as the regions of the dynamical influence
where all galaxy flows inside it are converging \cite{2}. 
High-density regions in the regions of the dynamical influence
correspond to conventional superclusters, and low-density regions
can be called as supercluster cocoons \cite{3}.
Supercluster cocoon boundaries follow 
the lowest density regions between superclusters  \cite{1}.

The aim of a detailed study of the distribution and properties of
galaxies and their systems in superclusters and in their
cocoons  is to clarify how the local and global
environmental effects combine in determining the evolution
and present-day dynamical state of superclusters and their components,
along with the star-formation history of galaxies in these structures.

In our study we analysed the structure and galaxy content of 
the supercluster SCl~A2142 at redshift $z \approx 0.09$, named after its richest cluster, A2142, 
and the cocoon region around it. 
We determined filaments in the cocoon, found the
connectivity of groups and clusters in and around the supercluster,
and the connectivity of the whole supercluster. 
The connectivity is defined 
as the number of filaments connected to a cluster, and it 
characterises the growth of cosmic structures \cite{4}. 
To understand the evolution of groups and galaxies in them, and 
possible cosmic web detachments \cite{5}
we studied  galaxies at various epochs of their star-formation
history.
Our study is the first study of a supercluster and its cocoon in which these
aspects are combined \cite{1}. 
We use the following cosmological parameters: the Hubble parameter $H_0=100~ 
h$ km~s$^{-1}$ Mpc$^{-1}$, matter density $\Omega_{\rm m} = 0.27$, and 
dark energy density $\Omega_{\Lambda} = 0.73$ 
\cite{6}.

\section{Supercluster, group, and filament data and supercluster definition}

We used data from supercluster, group, and filament catalogues, 
based on the Sloan Digital Sky Survey MAIN galaxy dataset, and 
available from the database of cosmology-related catalogues 
at http://cosmodb.to.ee/ \cite{7, 8, 9} and \cite{10}.
Data on galaxy properties are taken 
from the SDSS DR10 web page  
\footnote{http://skyserver.sdss3.org/dr10/en/help/browser/browser.aspx}.
Among many galaxy properties we used information
about the $D_n(4000)$ index of galaxies, which is 
calculated as the ratio of the average flux densities
in the band $4000 - 4100 \angstrom$ and $3850 - 3950 \angstrom$.
The $D_n(4000)$ index is correlated with the time passed 
from the most recent star formation event in a galaxy.
It can be used as proxy for the age of stellar populations of galaxies,
and star formation rates.

\begin{figure*}[ht]
\centering 
\hspace{2mm}  
\includegraphics[width=0.40\textwidth]{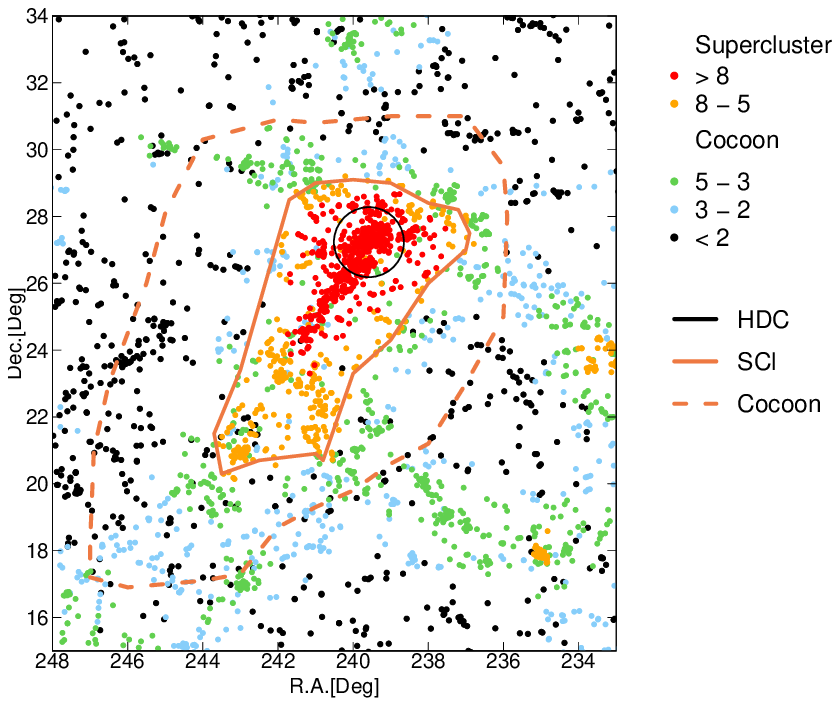}
 \hspace{2mm}  
\includegraphics[width=0.40\textwidth]{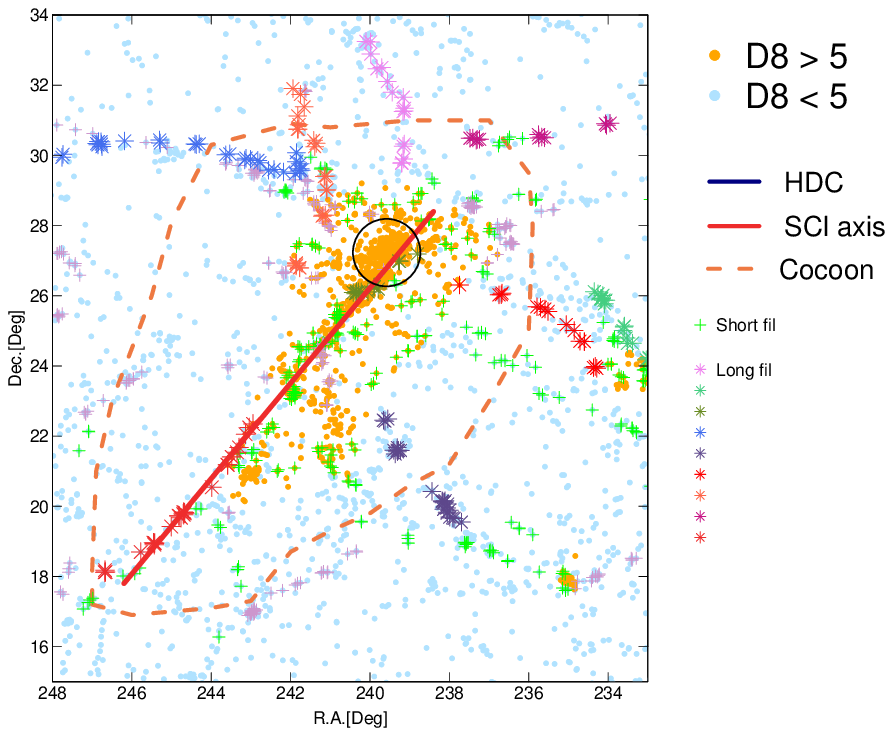}\\
\hspace{2mm}
\caption{Left: Distribution of galaxies in the supercluster SCl~A2142 region in the sky plane. 
Orange solid line indicate the boundaries of the supercluster, and
orange dashed line shows the cocoon boundaries. Black circle
shows the high-density core  of the supercluster. 
  Right:
Distribution of galaxies in SCl~A2142 and in its cocoon in the sky plane. 
Member galaxies of individual long filaments
are denoted with stars of different colours. 
Galaxies in short filaments are denoted with green and violet crosses,
Dark red line shows supercluster axis and straight filament at its extension. 
}
\label{EinastoMfig1.1}
\end{figure*}

We used the luminosity-density field to determine the 
supercluster SCl~A2142 and to find its cocoon boundaries. 
The supercluster centre is at the rich galaxy cluster A2142
at $R.A. =239.5$$^{\circ}$ and $Dec. = 27.3$$^{\circ}$.
The supercluster is defined as a volume around the cluster A2142 
with luminosity-densities of $D \geq 5.0$ (in units of mean luminosity-density, 
$\ell_{\mathrm{mean}}$ = 1.65$\cdot10^{-2}$ $\frac{10^{10} h^{-2} L_\odot}{(\vmh)^3}$,
in the luminosity-density field).
Cocoon boundaries are defined as minima in the density field around the supercluster.
We show the distribution of galaxies in SCl~A2142 and in its environment
in Fig.~\ref{EinastoMfig1.1} (left panel).
This figure shows that the supercluster has  a high-density core (hereafter HDC) with 
the radius approximately $5$~\Mpc. It is embedded in 
 an almost spherical main body of the supercluster with the 
luminosity-density $D \geq 8.0$. 
The outer parts of the supercluster
$5 \leq D < 8$ form its outskirts region.
SCl~A2142 has an almost straight tail,
and the total length of the supercluster is approximately $50$~\Mpc.  
The cocoon around the supercluster is elongated, with the maximum length of
about $80$~\Mpc.  

\section{Filaments and connectivity in the supercluster and in the cocoon}

Figure~\ref{EinastoMfig1.1} (right panel) shows long and short filaments
in SCl~A2142 and its cocoon. Filaments were identified using their member galaxies.
A galaxy was considered as a member of a filament if its distance from the 
filament axis was $D_{fil} \leq 0.5$~\Mpc.
We found that
six long filaments with length over $20$~\Mpc\ begin near the supercluster
main body. One long filament lies at the extension of the
supercluster tail. Therefore, the connectivity of the supercluster main body 
is $\pazocal{C} = 6$, and for the whole supercluster, it is $\pazocal{C} = 7$. 
The most interesting among long filaments which surround the supercluster
is a long filament, which lies at the extension of the supercluster tail.
It has an almost straight shape in the sky plane.
The total length of SCl~A2142 together with this filament
is approximately $75$~\Mpc, which makes it the longest straight structure
in the Universe described so far.

\section{Dynamical state of the supercluster and its high-density core}

\begin{figure*}[ht]
\centering 
\hspace{2mm}  
\includegraphics[width=0.40\textwidth]{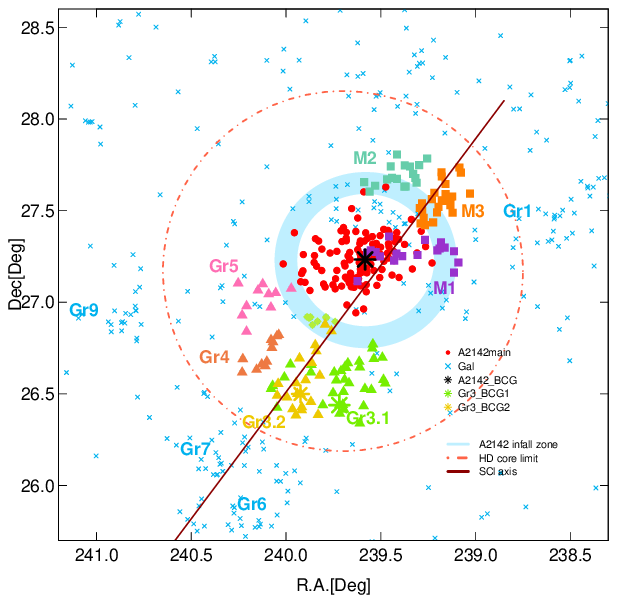}
 \hspace{2mm}  
\includegraphics[width=0.40\textwidth]{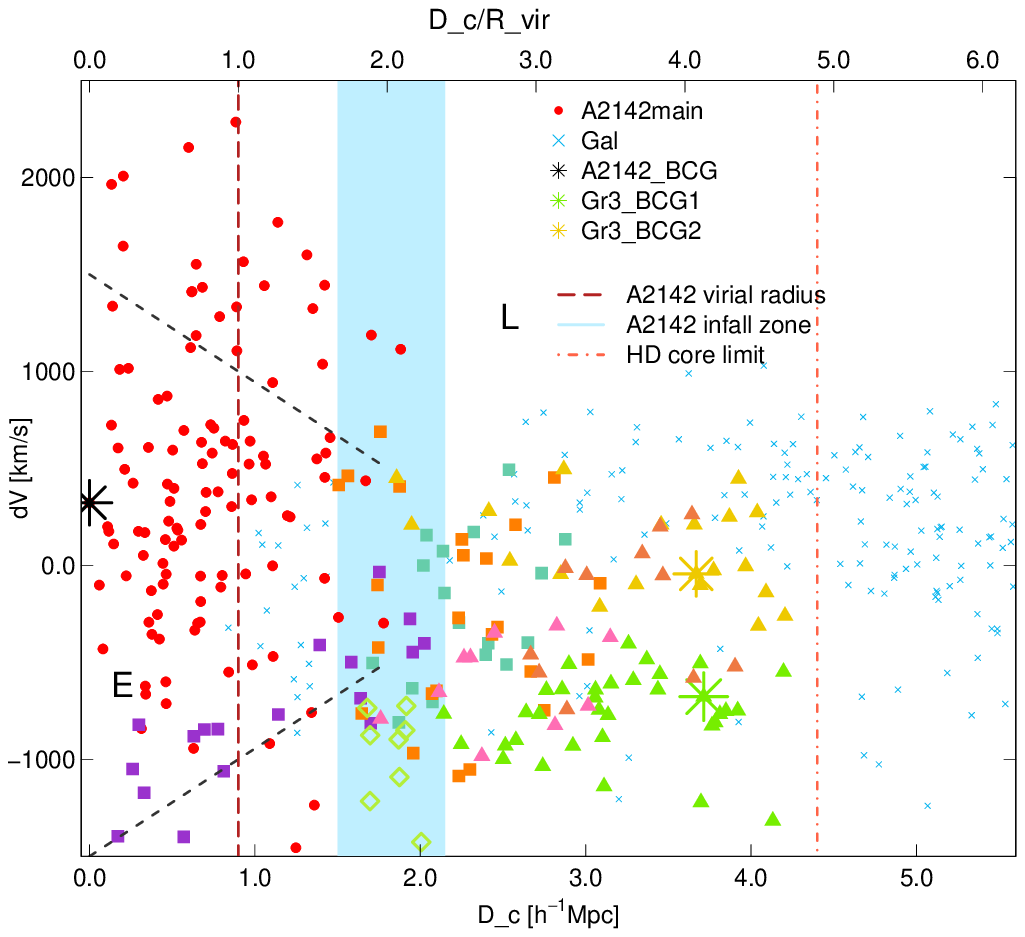}\\
\hspace{2mm}
\caption{Left: Distribution of galaxies of the supercluster SCl~A2142 HDC in the sky plane.  
The red symbols show galaxies with old stellar populations
($D_n(4000) \geq 1.55$), and the blue symbols denote galaxies with young 
stellar populations ($D_n(4000) < 1.55$).
  Right:
Velocity of galaxies with respect to the cluster mean velocity vs. 
projected distance from the centre of the A2142 cluster ($D_c$)
in the HDC (PPS diagram).
Dashed lines separate early and late infall regions. 
}
\label{EinastoMfig1.2}
\end{figure*}

\begin{figure*}[ht]
\centering 
\hspace{2mm}  
\includegraphics[width=0.40\textwidth]{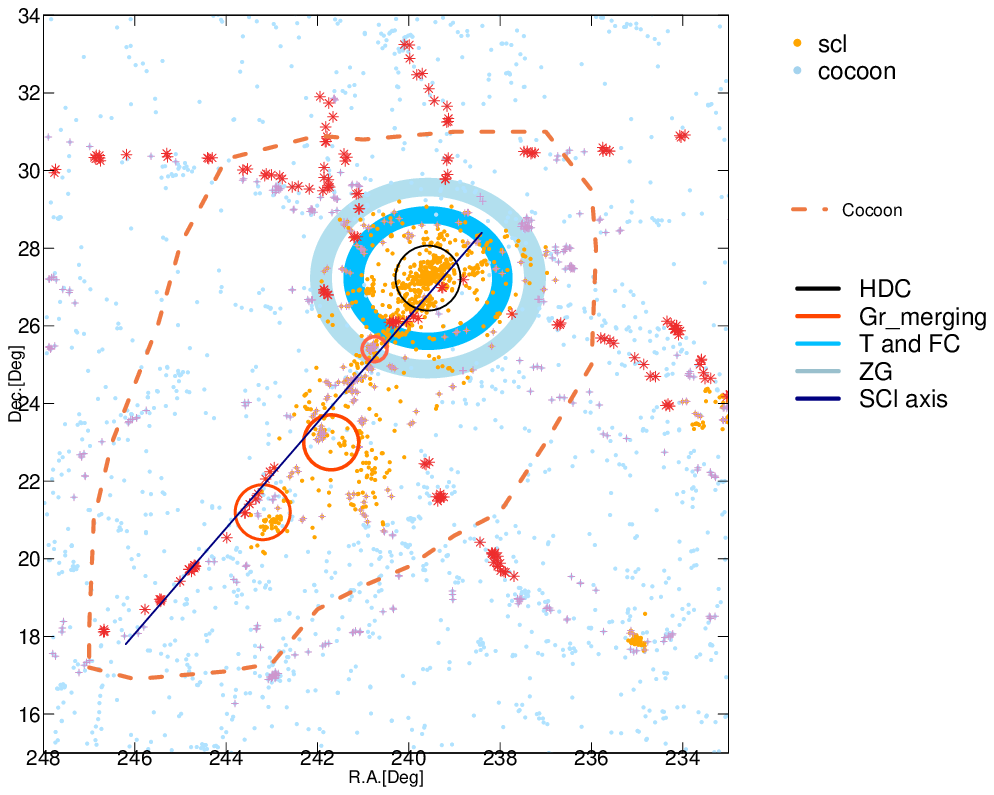}
 \hspace{2mm}  
\includegraphics[width=0.40\textwidth]{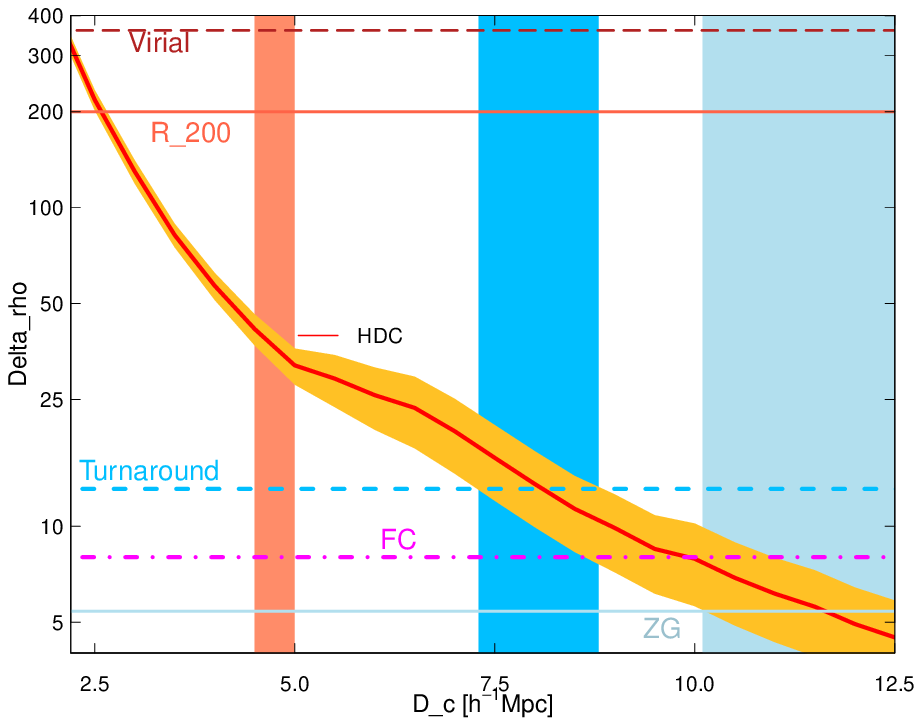}\\
\hspace{2mm}
\caption{Left: Distribution of galaxies in the sky plane in SCl~A2142 and in its cocoon. 
Dark red stars show galaxies in long filaments, and 
violet dots show galaxies in short filaments.
Orange circles mark the location of merging 
groups in the supercluster tail.
Blue stripe marks turnaround (T) region, 
and light blue circle shows borders of zero gravity (ZG)
region where long filaments are detached from the
supercluster. Future collapse (FC) region 
lies between them.
  Right:
Density contrast $\Delta\rho = \rho/\rho_{\mathrm{m}}$ versus
clustercentric distance $D_c$ for the SCl~A2142 main body (red line). 
Golden area shows error corridor
from mass errors. 
Characteristic density contrasts are denoted as follows: 
$\Delta\rho = 360$ (virial),
$\Delta\rho = 200$ ($r_{200}$), $\Delta\rho = 13.1$ (turnaround,
blue dashed line),
$\Delta\rho = 8.73$ (future collapse FC, magenta dash-dotted line), 
and $\Delta\rho = 5.41$
(zero gravity ZG, light blue solid line).
Tomato, blue, and light  blue vertical areas mark borders
of the HDC of the supercluster, 
turnaround region of the supercluster 
main body, and zero gravity region. 
}
\label{EinastoMfig2.1}
\end{figure*}

We start the analysis of the dynamical state of the supercluster
with the study of  its HDC. We show the sky distribution of 
galaxies in it, and present its projected phase space (PPS) 
diagram in Fig.~\ref{EinastoMfig1.2}.
In the PPS diagram  line-of-sight velocities of galaxies 
with respect  to the cluster mean velocity are plotted against 
the projected clustercentric distance (distance from the centre of A2142). 
Simulations show that in the PPS diagram 
galaxies at small clustercentric 
radii form an early infall (virialised) population E with 
infall times $\tau_{\mathrm{inf}} > 1$~Gyr, 
and galaxies 
at large clustercentric radii 
form late or ongoing infall population L with $\tau_{\mathrm{inf}} < 1$~Gyr.
Figure~\ref{EinastoMfig1.2} shows that the cluster A2142 is surrounded by substructures and groups,
and  the connectivity of A2142 is $\pazocal{C} = 6-7$.
They all lie in the infall zone of the cluster.  
Therefore, we may assume that the HDC is collapsing.
This assumption can be tested by the analysis of the mass distribution in the supercluster,
and by its comparison with the spherical collapse model, 
which is presented in Fig.~\ref{EinastoMfig2.1}.
To calculate the radial mass distribution in the supercluster for finding density
contrast we used dynamical masses of groups from the group catalogue.

The spherical collapse model describes the evolution 
and dynamical state of a spherical perturbation in an expanding universe 
using various epochs, each of which has  characteristic
density contrast (see \cite{11} for details and references).
The first epoch  is the 'turnaround', with
a density contrast of $\Delta\rho_{T} = 13.1$, 
defined as the epoch 
at which the collapse begins.
Overdensity regions in which density contrast is not sufficient
to collapse at present may 
collapse in the future, if their current overdensity is
$\Delta\rho_{FC} = 8.73$ (FC denotes future collapse).
The density contrast $\Delta\rho_{ZG} = 5.41$ corresponds to a so-called zero 
gravity radius (ZG) where  gravitation equals  expansion.
It borders the region which may stay gravitationally bound. 
  
The HDC of the supercluster has 
density contrast
at its borders $\Delta\rho \approx 30$.
This suggests   
that the HDC has passed turnaround, and continues contracting. 
The radius of the turnaround region
of the supercluster main body, $R_T \approx 7 - 9$~\Mpc. 
Turnaround region is populated by poor groups and single
galaxies in short filaments.  Long filaments do not reach 
the inner regions of the supercluster, and this may be related to the 
collapse of the supercluster main body, which have destroyed the structures in this 
region. The outer parts of the supercluster main body with radius 
$R_{FC} \approx 9$~\Mpc\ will collapse in the future. 
Zero gravity region with a radius of 
$R_{ZG} \approx 10 - 13$~\Mpc\ approximately surrounds the supercluster
main body. 
This means that regions at larger distances from the supercluster centre
will not become gravitationally bound. Groups in the supercluster tail are merging, and
will probably form at least three separate systems in the future.
We note that these density contrasts correspond to cosmological model with 
matter density $\Omega_{\rm m} = 0.27$. If $\Omega_{\rm m} \approx 0.3$,
then density contrast at runaround is  $\Delta\rho_{T} = 12.2$ \cite{11},
and the turnaround region of SCl~A2142 is larger.

The characteristic  radii 
in Fig.~\ref{EinastoMfig2.1} were found using mass
estimates  of groups. 
These radii  are in a good accordance with those
at which long  filaments are detached. 
This coincidence supports our interpretation that long filaments are detached 
because of the collapse.
This suggests that the detailed study of filaments in and around superclusters
together with the study of the dynamical state of superclusters
can be used as a cosmological test. 

\section{Galaxy populations in the supercluster and in the cocoon}

We analysed the spatial distribution of galaxies with different star 
formation histories in the supercluster and in cocoon, and
found several interesting trends which are described in \cite{1}. 
Next we focus on the distribution of galaxies with very old stellar populations (VO),
defined as having $D_n(4000)$ index values $D_n(4000) \geq 1.75$.
The value $D_n(4000) = 1.75$ corresponds to 
the mean age of about $4$~Gyr \cite{12}. 

\begin{figure*}[ht]
\centering 
\hspace{2mm}  
\includegraphics[width=0.40\textwidth]{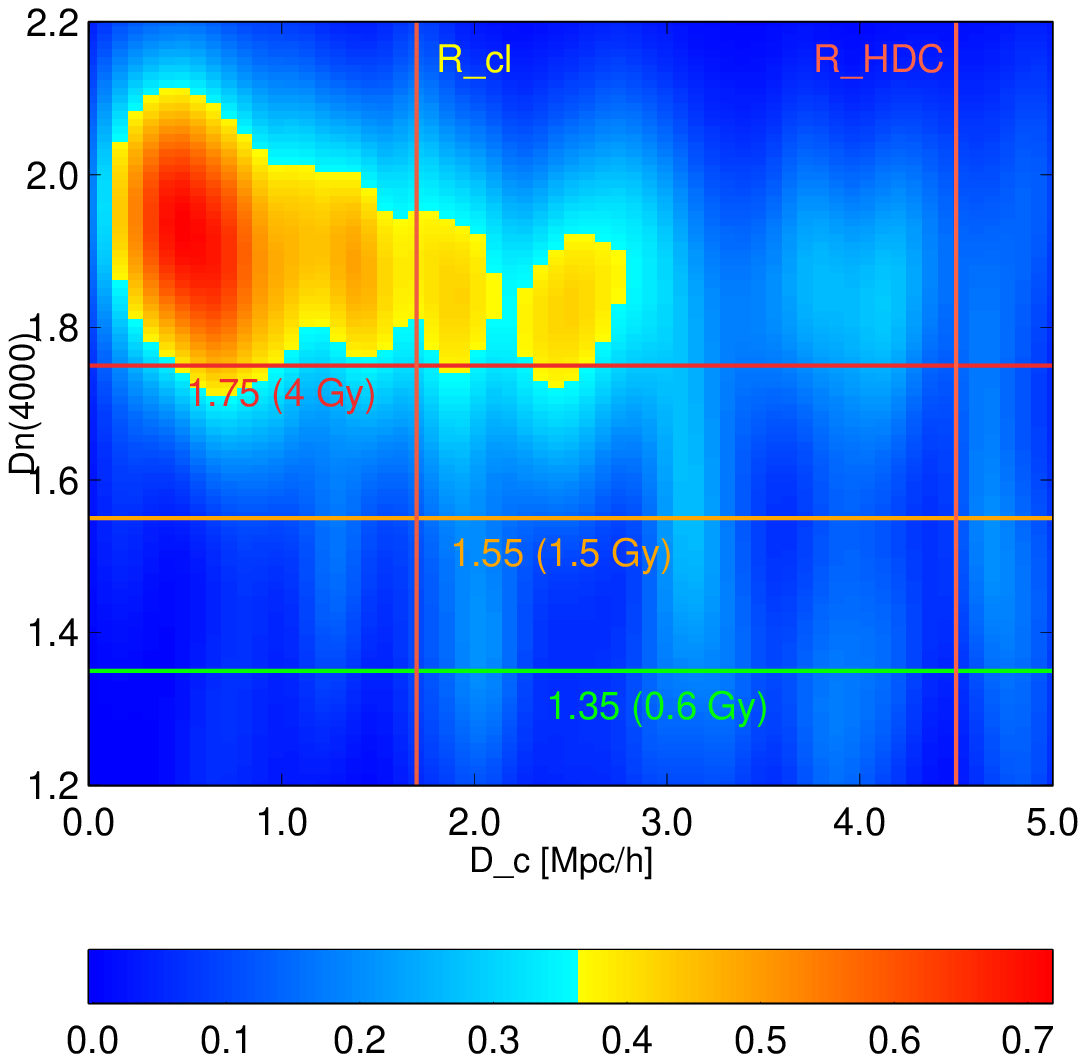}
 \hspace{2mm}  
\includegraphics[width=0.44\textwidth]{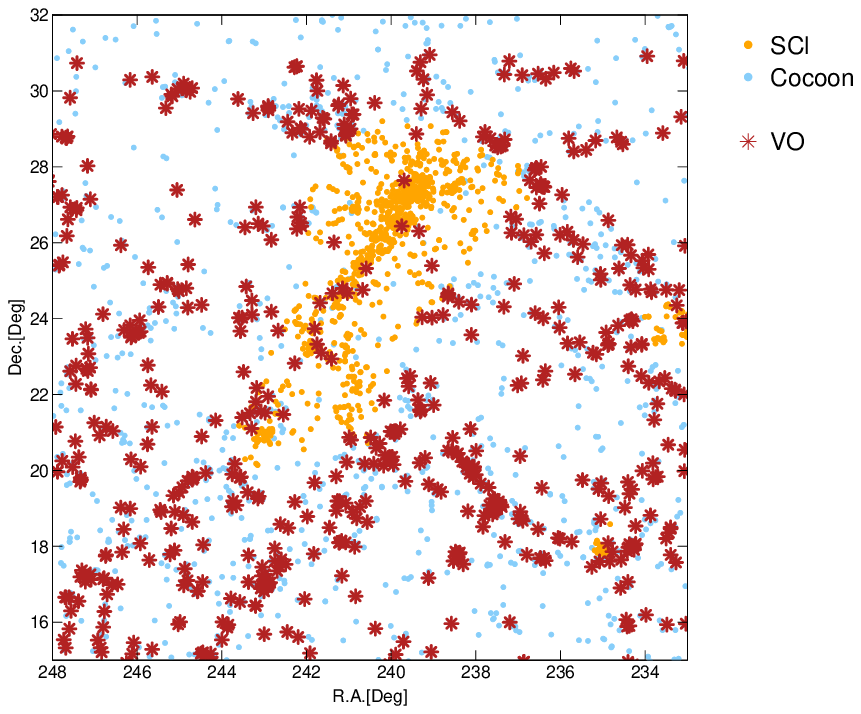}\\
\hspace{2mm}
\caption{Left: $D_n(4000)$ index versus clustercentric distance  $D_c$
for galaxies in the cluster A2142 and in the HDC. 
Colours show the density of points at the given location in the plot.
Horizontal lines show $D_n(4000)$ index values
$D_n(4000) = 1.75$ (VO galaxies), 
$D_n(4000) = 1.55$ (limit between quenched and star forming galaxies), and $D_n(4000) = 1.35$ 
(galaxies with very young stellar populations). 
  Right:
Distribution of galaxies in SCl~A2142 and in its cocoon in the sky plane. 
Dark red stars show VO galaxies with $D_n(4000) \geq 1.75$.
}
\label{EinastoMfig2.2}
\end{figure*}
 
In Fig.~\ref{EinastoMfig2.2} (left panel) we show the clustercentric distribution of VO galaxies
in the HDC of SCl~A2142. This figure shows that the central, virialised part of 
the cluster A2142 (the early infall region in the PPS diagram 
in Fig.~\ref{EinastoMfig1.2}) is populated
with VO galaxies, and there are some VO galaxies also in the infalling structures.
This is expected - galaxies 
which falled into the cluster during its formation, have been quenched there under
influence of various processes \cite{1, 13}. Therefore, 
we may expect that very dense environments are typical environments for VO galaxies.
Figure~\ref{EinastoMfig2.2} (right panel) shows the sky distribution of VO galaxies in the cocoon.
Surprisingly, these galaxies can be found everywhere in the cocoon, in various structures
from groups and filaments to single galaxies. 
The richness of their host groups vary from single galaxies to the richest 
cluster in our study, A2142, with  several orders of host group mass range, from
$10^{13}h^{-1}M_\odot - 10^{15}h^{-1}M_\odot$ \cite{1}. 
In filaments, they lie in the central parts,
while outer parts are populated by star-forming galaxies \cite{1}. 
Their properties are nearly independent of the global environmental 
density and, thus, the overall conditions for these galaxies to form and evolve 
had to be similar 
(or leading to similar galaxies) both in high- and low-global-density environments. 
This is unexpected owing to the very different conditions
in rich and  poor groups.

\section{Summary}

In our study of the supercluster SCl~A2142 and its cocoon
we determined the boundaries of the supercluster SCl~A2142 cocoon using the luminosity-density 
field and studied the galaxy, group, and filament content therein.
Our study showed that the connectivity of the supercluster
and the galaxy content of groups and filaments are related to the 
evolution of the supercluster and its cocoon. Our results
can be summarised as follows.

\begin{itemize}
\item[1)]
The supercluster A2142, together with the long filament connected to it, forms
the longest straight structure in the Universe detected so far, with a 
length of approximately $75$~\Mpc.
The length of the supercluster itself is $\approx 50$~\Mpc, and the size of cocoon
is $\approx 80$~\Mpc.
\item[2)]
The supercluster  main body
is collapsing and six long filaments which surround it
are detached from it  at the turnaround region.
The whole supercluster 
has connectivity $\pazocal{C} = 7$. 
\item[3)]
Among various trends in the properties of galaxies and groups
related to the 
local (group and cluster) and global (supercluster and cocoon) environment
 we found that galaxies with the oldest stellar populations 
can be found in extremely different environments from the
centre of the richest cluster in the superclusters to single galaxies in 
cocoon filaments.  
\end{itemize}

The study of 
the supercluster SCl~A2142, galaxies and galaxy groups in it
and in its environment shows how the supercluster and structures around it evolve, 
making such type of studies a valuable tool
in cosmology to study the evolution of the cosmic web. 
The supercluster has 
a long, straight filament as the extension of the supercluster,
forming the longest straight structure detected so far.
The presence of such structures may be a challenge for the theories
of the structure formation in the Universe. 
 


\bigskip

I thank my co-authors Boris Deshev, Peeter Tenjes, Pekka Hein\"am\"aki,
Elmo Tempel, Lauri Juhan Liivam\"agi, Jaan Einasto, Heidi Lietzen, Taavi Tuvikene,
and Gayoung Chon for fruitful and enjoyable collaboration.
We are pleased to thank the SDSS Team for the publicly available data
releases. The SDSS website is \texttt{http://www.sdss.org/}.
We applied in this study R statistical environment.

\section*{Funding}
The present study was supported by the ETAG projects 
IUT26-2, IUT40-2,  PUT1627, by the European Structural Funds
grant for the Centre of Excellence "Dark Matter in (Astro)particle Physics and
Cosmology" (TK133), and by MOBTP86.


\end{document}